\documentclass[ajl]{emulateapj}

\usepackage{graphicx,epsfig,natbib,color}

\usepackage[usenames,dvipsnames,svgnames,table]{xcolor}
\usepackage[colorlinks=true,
           linkcolor=blue,
           urlcolor=blue,
           citecolor=blue]{hyperref}

\newcommand{\kms}{km~s$^{-1}$}

\begin{document}

\title{PARTIAL REFLECTION AND TRAPPING OF A FAST-MODE WAVE IN SOLAR CORONAL ARCADE LOOPS}
\author{PANKAJ KUMAR\altaffilmark{1,2}, D.E. INNES\altaffilmark{2}}
\affil{$^1$Korea Astronomy and Space Science Institute (KASI), Daejeon, 305-348, Republic of Korea}
\affil{$^2$Max-Planck Institut f\"{u}r Sonnensystemforschung, 37191 Katlenburg-Lindau, Germany}
\email{pankaj@kasi.re.kr}

\begin{abstract}
We report on the first direct observation of a fast-mode wave propagating
along and perpendicular to cool (171~\AA) arcade loops observed by the {\it
Solar Dynamics Observatory/Atmospheric Imaging Assembly} (AIA). The wave was
associated with an impulsive/compact flare, near the edge of a sunspot. The
EUV wavefront expanded radially outward from the flare center and decelerated
 in the corona from 1060-760 km s$^{-1}$ within $\sim$3-4 minute.
Part of the EUV
wave propagated along a large-scale arcade of cool loops and was partially
reflected back to the flare site.  The phase speed of the wave was about 1450
km s$^{-1}$, which is interpreted as a fast-mode wave. A second
overlying loop arcade, orientated perpendicular to the cool arcade, is heated and
becomes visible in the AIA hot channels. These hot loops sway in time with the
EUV wave, as it propagated to and fro along the lower loop arcade. We suggest that an impulsive energy
release
at one of the footpoints of the arcade
loops causes the onset of an EUV shock wave that propagates along and
perpendicular to the magnetic field.
\end{abstract}

\section{INTRODUCTION}
The study of Magnetohydrodynamic (MHD) waves in the solar atmosphere is very
important because they provide an indirect way to probe the solar corona
via coronal seismology and may play an important role in coronal heating
\citep{uchida1970,roberts1984,nakariakov2005,banerjee2007,demoortel2012}.

Large-scale coronal waves are generally observed during solar eruptions.
These waves were discovered by the {\it Extreme-ultraviolet Imaging Telescope}
(EIT) onboard the {\it Solar and Heliospheric Observatory} (SOHO), and
are sometimes referred to
 as EIT waves \citep{thompson1999}. The nature of these waves is still under
debate. Now most of the observations from SDO/AIA and STEREO support their
wave nature. These waves can be impulsively generated either by thermal
pressure produced by the flare or magnetic pressure of the CME piston
\citep{vrsnak2008}. Using STEREO observations, \citet{veronig2010} reported
on a dome-shaped EUV wave (speed $\sim$650 km s$^{-1}$) and interpreted it as a
weakly shocked fast-mode wave. At present, there are several
observational studies of propagating fast-mode waves using SDO/AIA images (e.g.,
\citealt{chen2011},\citealt{ma2011}, \citealt{pats2012}, \citet{nitta2014},
\citealt{liu2014} and references cited therein). Furthermore, transverse kink
oscillations of the loops/filaments (located away from the flare) are also
observed when a fast MHD wave (i.e., global EUV wave) interacts with these structures
\citep{asai2012, kumar2013b}.

Recently, fast-mode wave trains have been discovered (during solar
flares/eruptions) with SDO/AIA, propagating along funnel shaped loops with
the speed of $\sim$1000-2000 km s$^{-1}$ \citep{liu2011, shen2012,
kumar2013a}. Using Hinode/SOT observations, \citet{ofman2008} reported the
first observations of transverse waves in coronal multi-threaded loops with
cool plasma ejected from the chromosphere flowing along the threads. They
found that the waves were nearly standing (fundamental) kink modes in some of
the treads with a phase speed of $\sim$1250 km s$^{-1}$, whereas the dynamics
of other threads was consistent with propagating fast magnetoacoustic waves.
Apart from the fast-mode wave, reflecting slow-mode wave (in arcade loops)
 has been recently discovered in the AIA hot channels (131 and 94 \AA~) \citep{kumar2013},
which was consistent with the SOHO/SUMER Doppler-shift oscillations (T$>$6 MK) 
\citep{kliem2002,wang2002,wang2003a,wang2003b}.  
However, fast-mode wave propagation along closed arcade loops, its
partial reflection and trapping has not been reported so far.


In this letter, we report the first direct observation of fast-mode wave
propagation (along and across arcades of loops) and its partial reflection in
cool (AIA 171 \AA) loops. The EUV wave caused transverse oscillation, seen in the AIA 131 and 94~\AA\ channels, of
loops orientated perpendicular and above the cool loop arcade.
The EUV wave was observed during an
impulsive/compact flare that occurred at one of the footpoints of the
cool loop system on 6 March 2014.  In section 2, we
present the observations and in the last section, we discuss and
summarize the results.
\begin{figure*}
\centering{
\includegraphics[width=8cm]{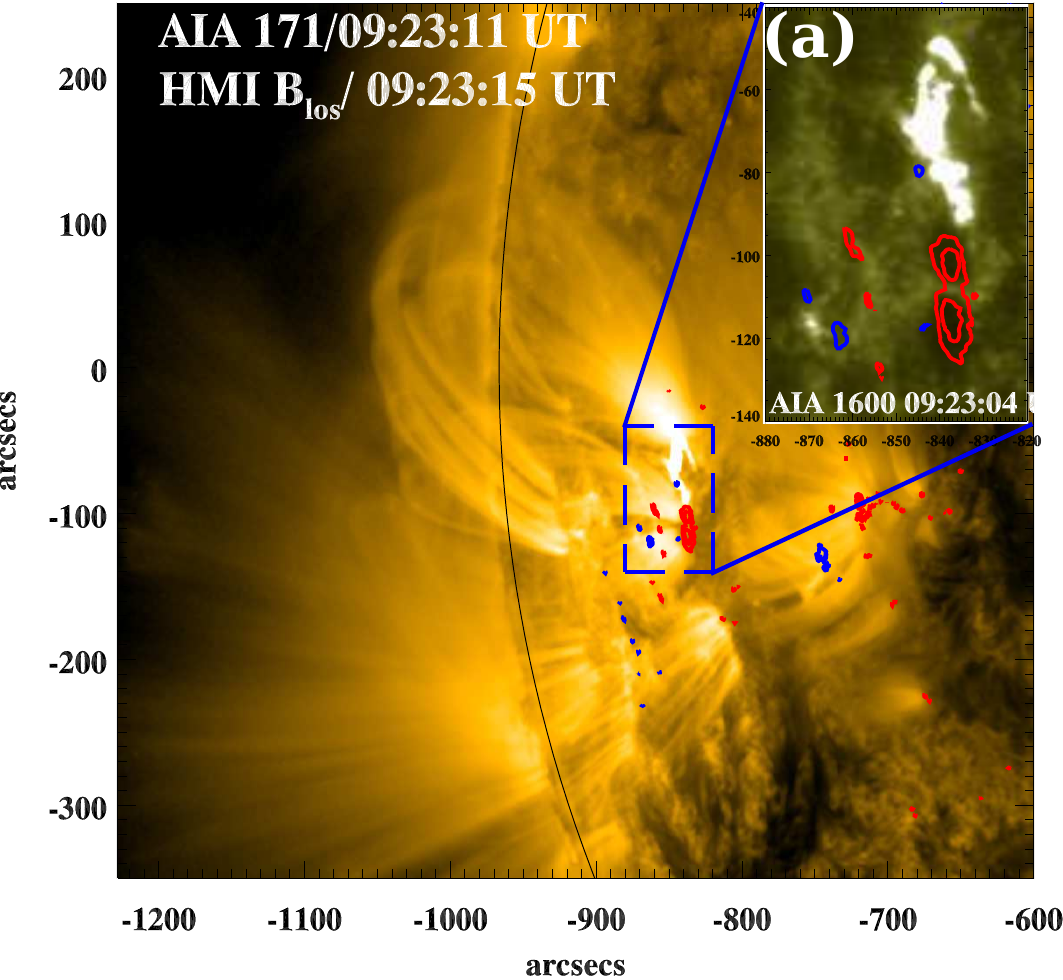}
\includegraphics[width=8cm]{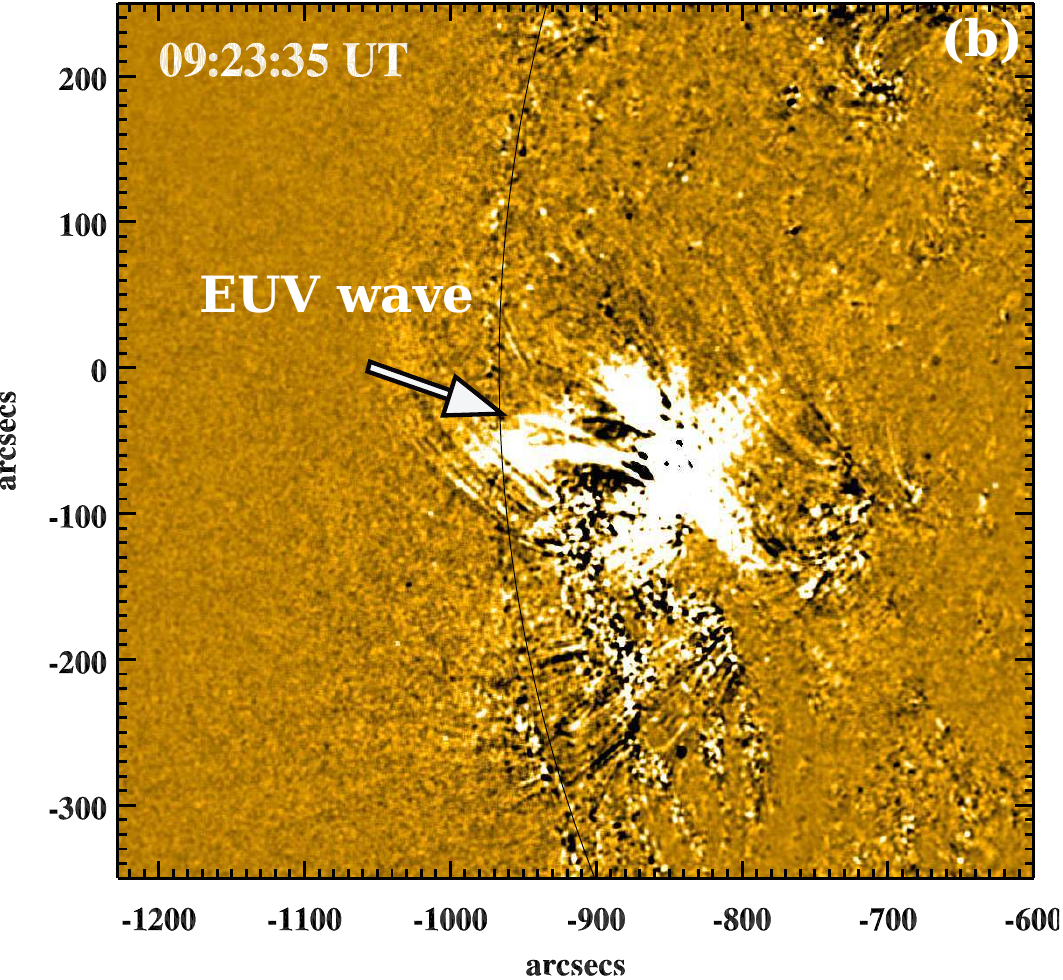}

\includegraphics[width=8cm]{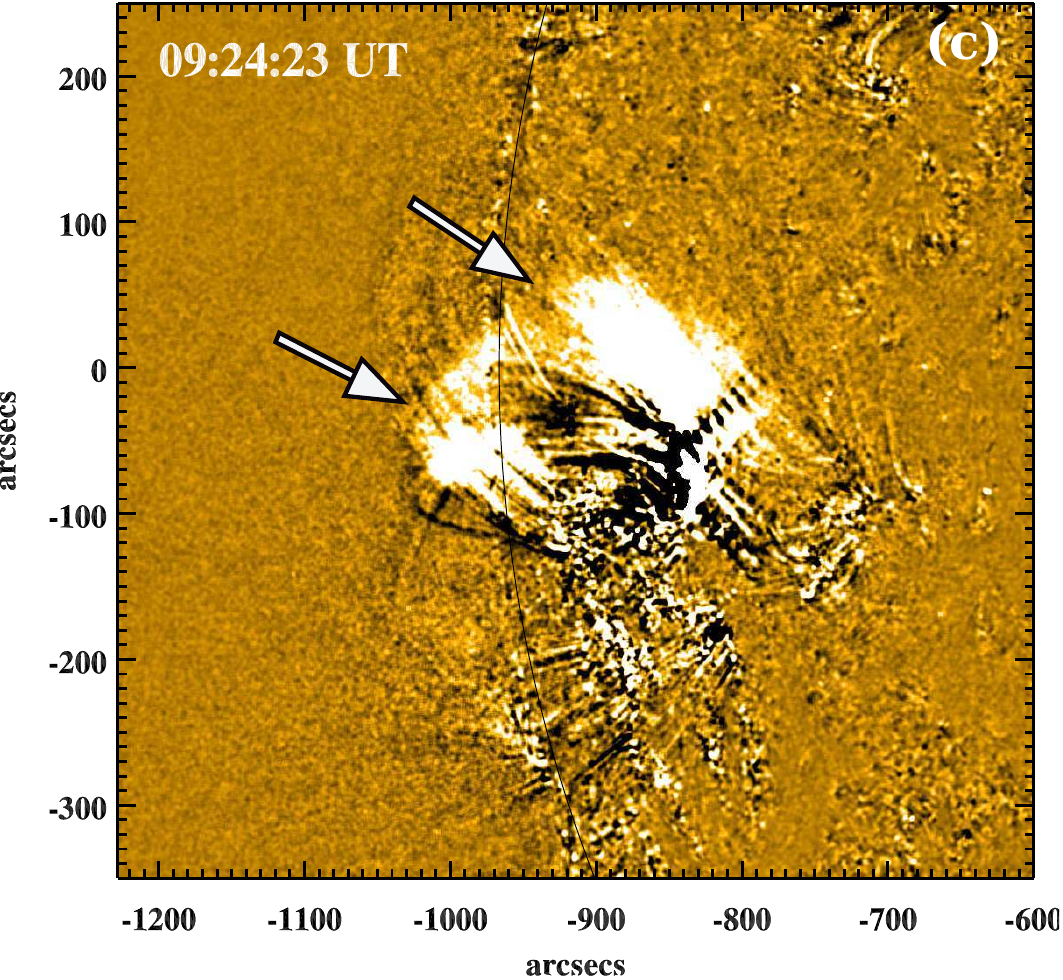}
\includegraphics[width=8cm]{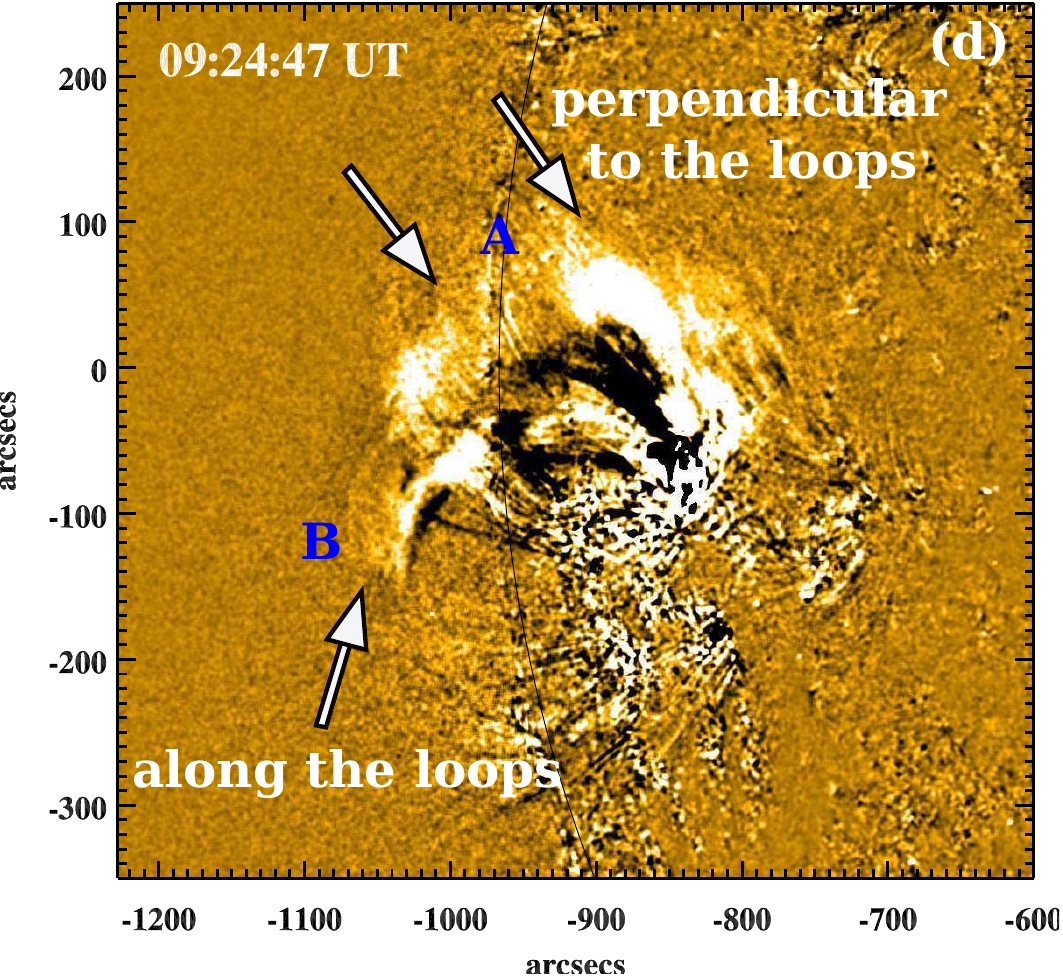}

\includegraphics[width=8cm]{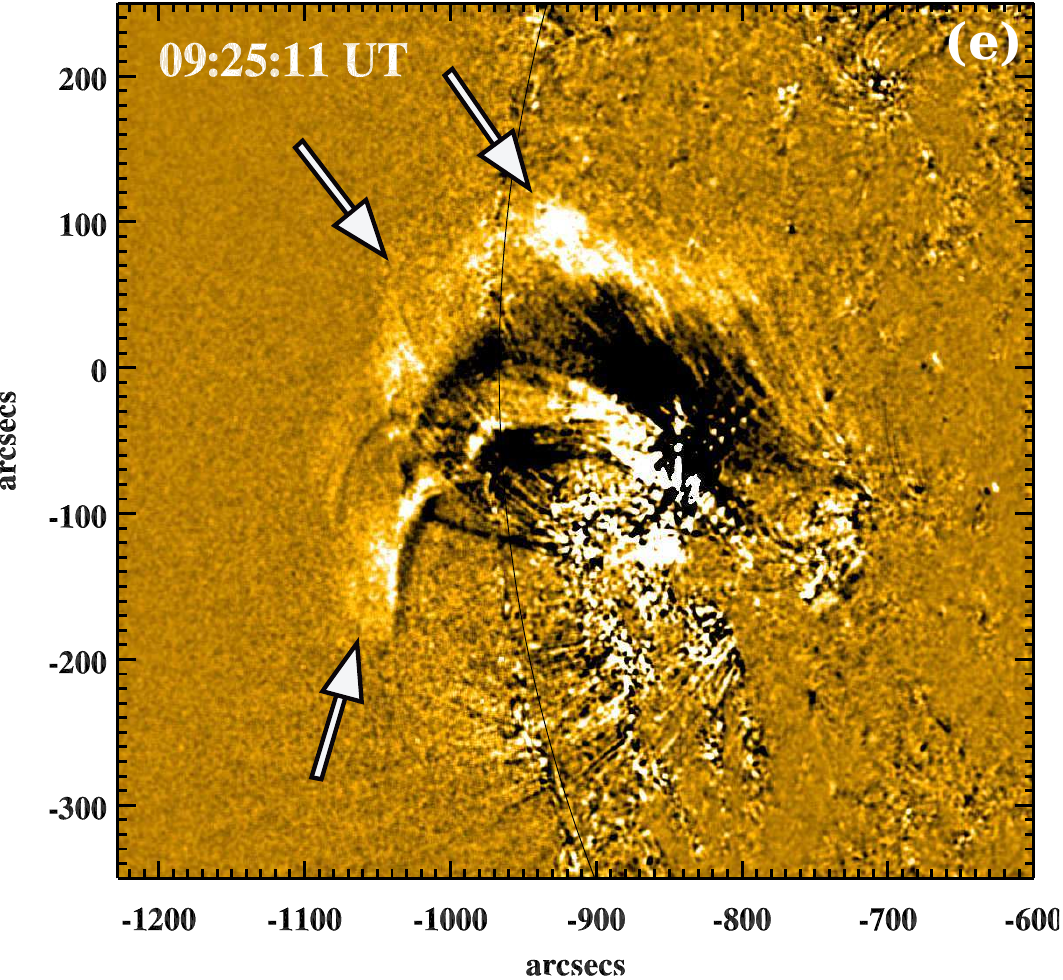}
\includegraphics[width=8cm]{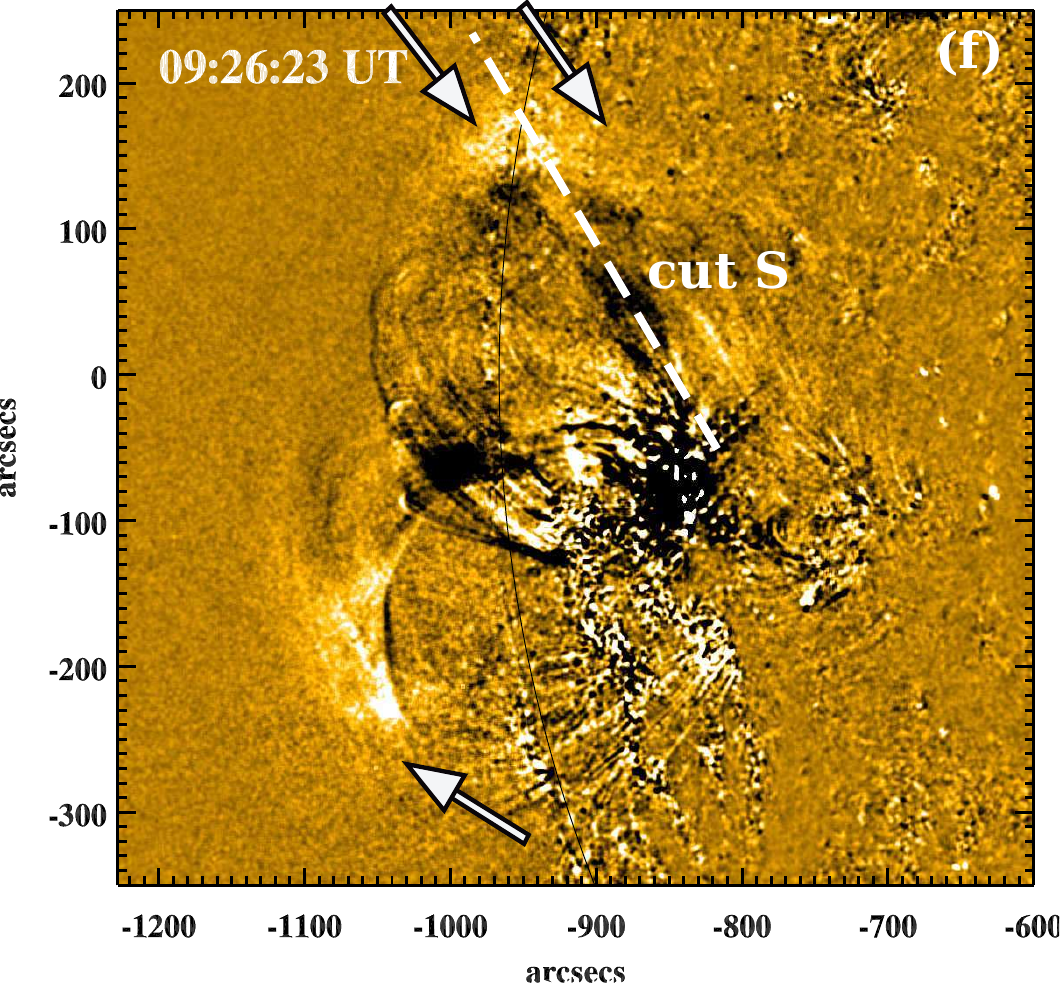}
}
\caption{(a) HMI magnetogram contours of positive (red) and negative (blue) polarities over 171 \AA~ intensity image. 
The contours levels are $\pm$400, $\pm$800 G.
Inset shows the flaring region in 1600 \AA~ channel, which is overlaid by HMI magnetogram contours.
(b-f) AIA 171 \AA~ running difference images showing the propagating EUV wave moving along and perpendicular to the loop system. (Animation is available.)}
\label{171_rd}
\end{figure*}
\begin{figure*}
\centering{
\includegraphics[width=13cm]{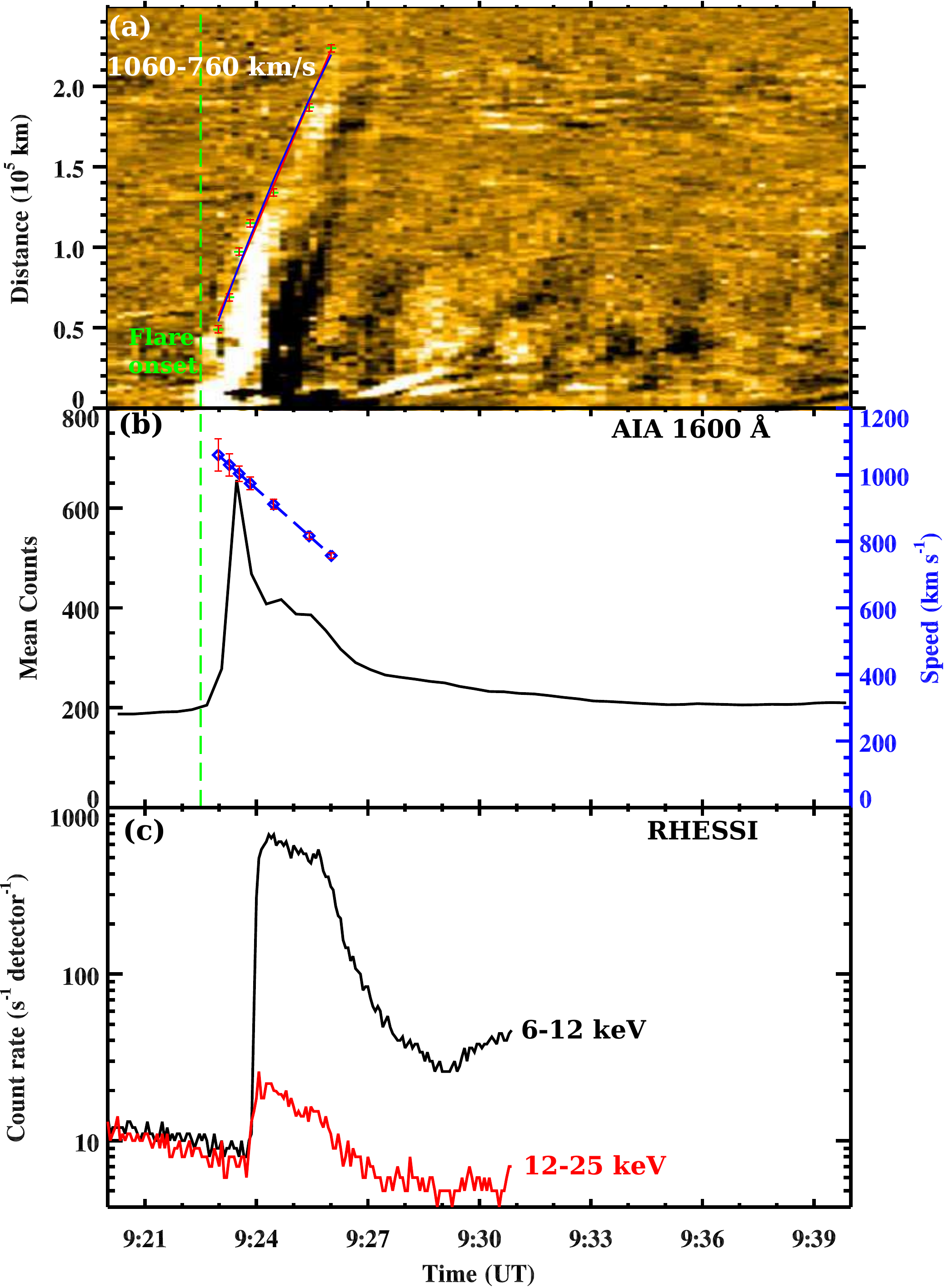}
}
\caption{(a) Distance-time plot of the intensity distribution along the slice `S' using 171 \AA~ running difference images. The linear fit to the data point is shown by the red curve. (b) Mean intensity profile of the flaring region, extracted using a box region at the flare site in 1600 \AA~ images. Speed profile of the EUV wave derived from a second order polynomial fitting to the wavefront in the distance-time plot (blue curve). (c) RHESSI X-ray flux profiles. Green vertical line shows the start time of the flare and associated EUV wave.}
\label{stack}
\end{figure*}
\section{OBSERVATIONS AND RESULTS}
The {\it Atmospheric Image Assembly} (AIA; \citealt{lemen2012}) onboard the
{\it Solar Dynamics Observatory} (SDO) records
full disk images of the Sun (field-of-view $\sim$1.3 R$_\odot$) with a
spatial resolution of 1.5$\arcsec$ (0.6$\arcsec$ pixel$^{-1}$) and a cadence
of 12~s. For the present
study, we utilized 171~\AA\ (Fe IX, with formation temperature $T\approx$0.7
MK), 94~\AA\ (Fe XVIII, $T\approx$6.3 MK), and 1600~\AA\ (C IV + cont.,
$T\approx$0.01 MK) images.

AR NOAA 11198 ($\beta$ magnetic configuration,
S07E64) was located on the eastern limb on 6 March 2014. The EUV wave,
reported here, was associated with an impulsive/compact flare that started
(in this AR) at $\sim$09:23 UT, maximized at $\sim$09:25 and ended at
$\sim$09:30~UT.

\subsection{Observation of the EUV wave}
The EUV wave is best observed in the AIA 171~\AA\ channel. We utilized AIA
171~\AA\ running difference images ($\Delta$t=1 min) to study the kinematics
and propagation characteristics of the EUV wave. Figure \ref{171_rd} displays
 selected AIA 171 \AA~ running difference images. We overlaid HMI
magnetogram \citep{schou2012} contours of positive (red) and negative (blue) polarities
on
the
AIA 171~\AA\ intensity image in the first panel (a), to show the magnetic
field configuration of the flare site. The inset shows the AIA 1600 \AA~ image,
exhibiting a closer view of the flare site. The brightening of a flare
ribbon starts ($\sim$09:23~UT) at the northern edge of a positive polarity
sunspot. Figures \ref{171_rd}(b) and (c) show the outward movement of the EUV
disturbance/wavefront from the flare center. Figure \ref{171_rd}(d) clearly
shows the EUV wavefronts propagating perpendicular and parallel to the
active region loops (09:24:27 UT). These fronts are marked by `A' and `B',
respectively. Later, we noticed that the wave reflected (front B) from the
other footpoint of the arcade loops. When the returning wave reached point `B', the
loops appeared to expand upwards, generating motion both outward into the corona
and back to the flare site
(see AIA 171 \AA~ running difference movie).

The EUV
disturbance was also observed in the AIA 193 and 211 \AA~ channels, which provide
better contrast for large-scale EUV waves in the solar corona (see online
movie). The circular EUV wavefront shows outward expansion from the flare
site.
The radially outward moving front (A) could be tracked until about 09:27 UT
in the AIA field of view.
To determine the kinematics of the outward moving EUV wavefront (A), we used
a slice cut S (Figure \ref{171_rd}f) to create a distance-time plot. Figure
\ref{stack}a displays the stack plot of running difference intensity along
the slice S. The wavefront is clearly observed from $\sim$09:23 UT to 09:26
UT. We estimated the mean speed of the wavefront, using a linear fit to the
visually selected data points (red curve), to be $\sim$910$\pm$10 km
s$^{-1}$. We assumed 5 pixel uncertainty in the identification of the EUV
wavefront.
To deduce the speed profile (range) of the wavefront,
 we fitted a second order polynomial function (blue curve) to the observed
 data points and estimated the speed. 
Figure \ref{stack}(b) shows the speed profile of the wave.
 The EUV wave decelerates from
 $\sim$1060 to 760 km s$^{-1}$ in the 3-4 minute interval.

To examine the relationship between the EUV wave and the flare, we plotted
the AIA 1600 \AA~ mean flux of the flare region as shown in Figure
\ref{171_rd}a. 
Figure \ref{stack}(b-c) shows the AIA 1600~\AA\
flux profile and  RHESSI X-ray flux profiles in the 6-12 and 12-25 keV energy channels. It is
evident that the onset of the EUV wave and the rising of the 1600~\AA\ flux evolve
simultaneously. Therefore, the trigger of the EUV wave is closely related with
the flare energy release. There was no plasmoid eruption during the flare
onset in any of the AIA channels (i.e., AIA 304, 131, 94 etc.).

\begin{figure*}
\centering{
\includegraphics[width=8.0cm]{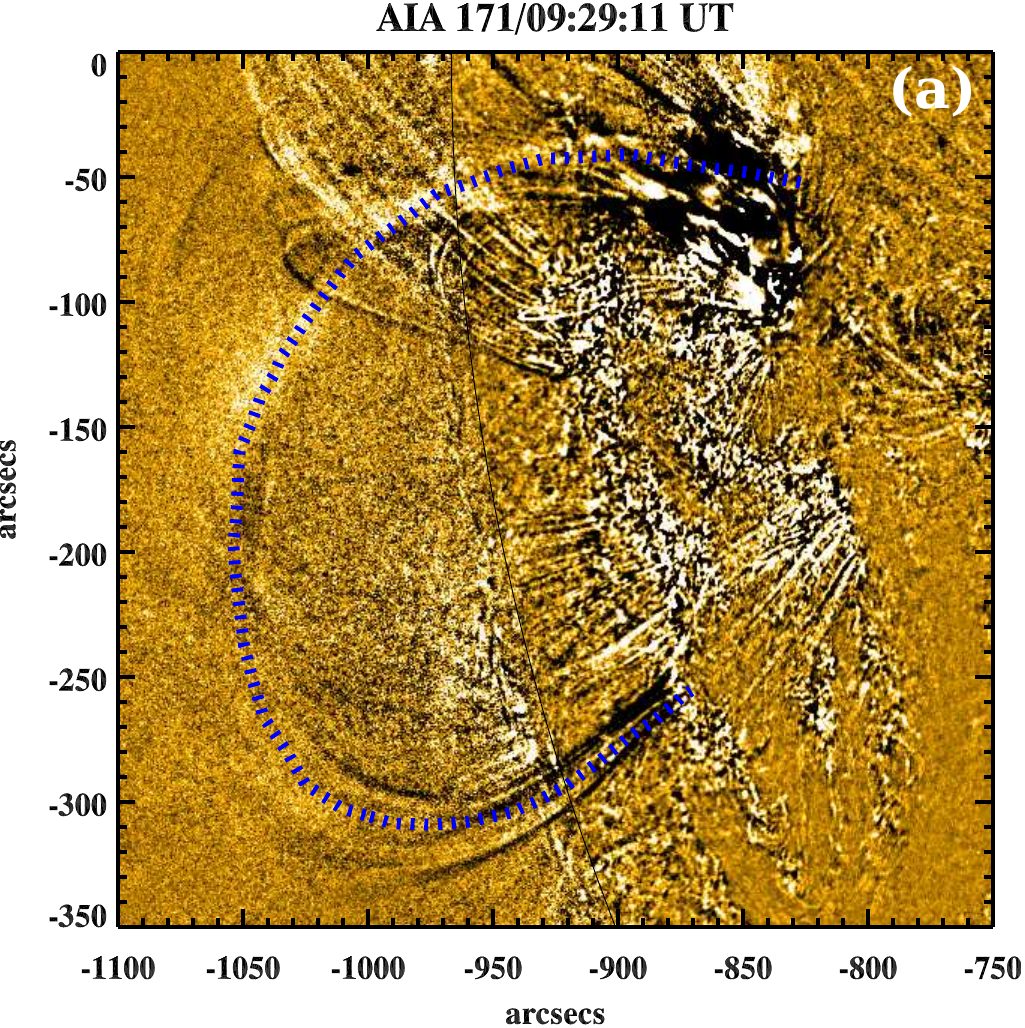}
\includegraphics[width=4.0cm]{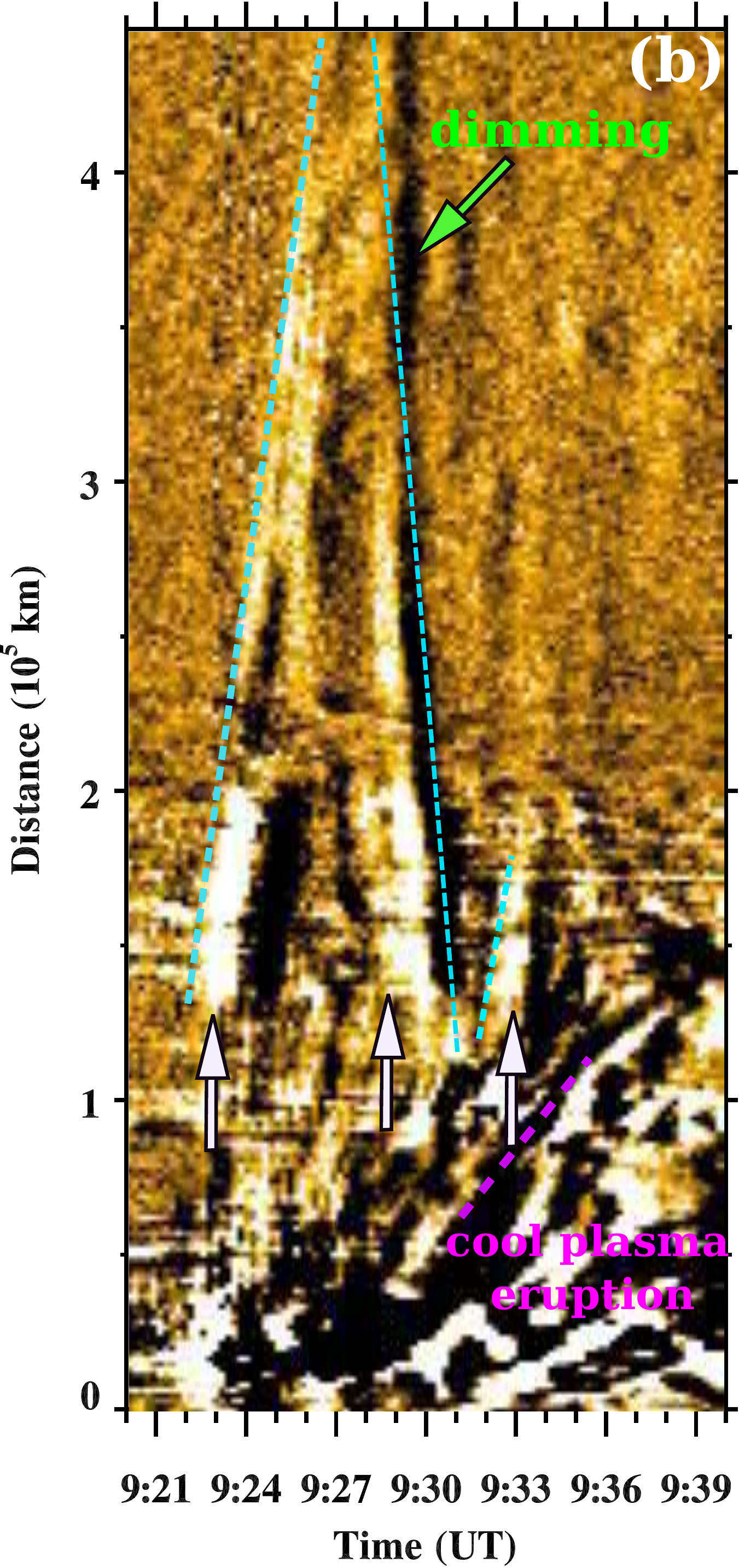}

\includegraphics[width=7.5cm]{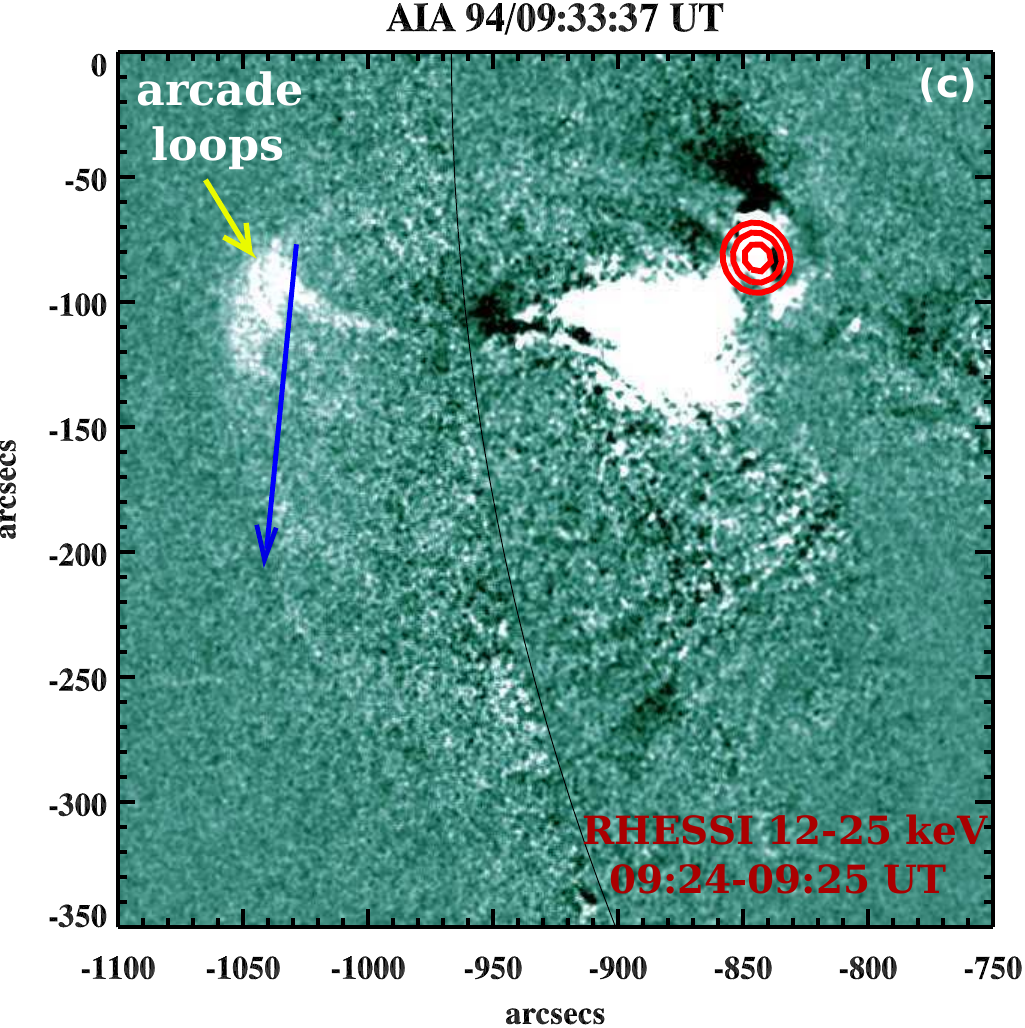}
\includegraphics[width=7.5cm]{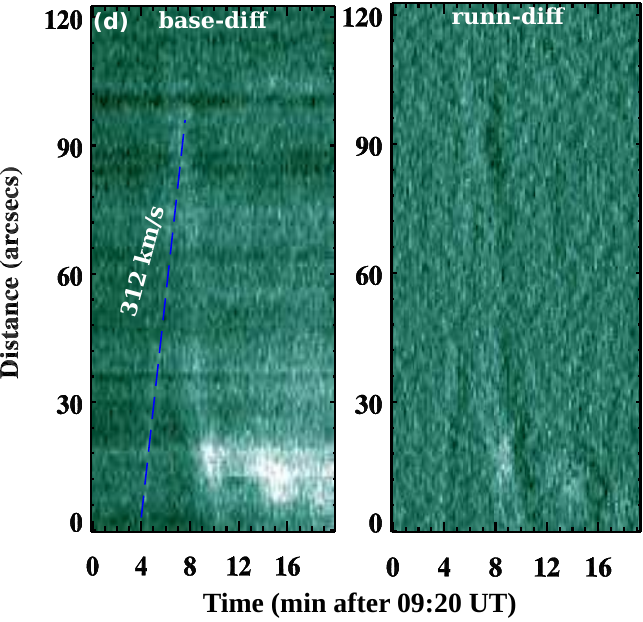}

\includegraphics[width=7.0cm]{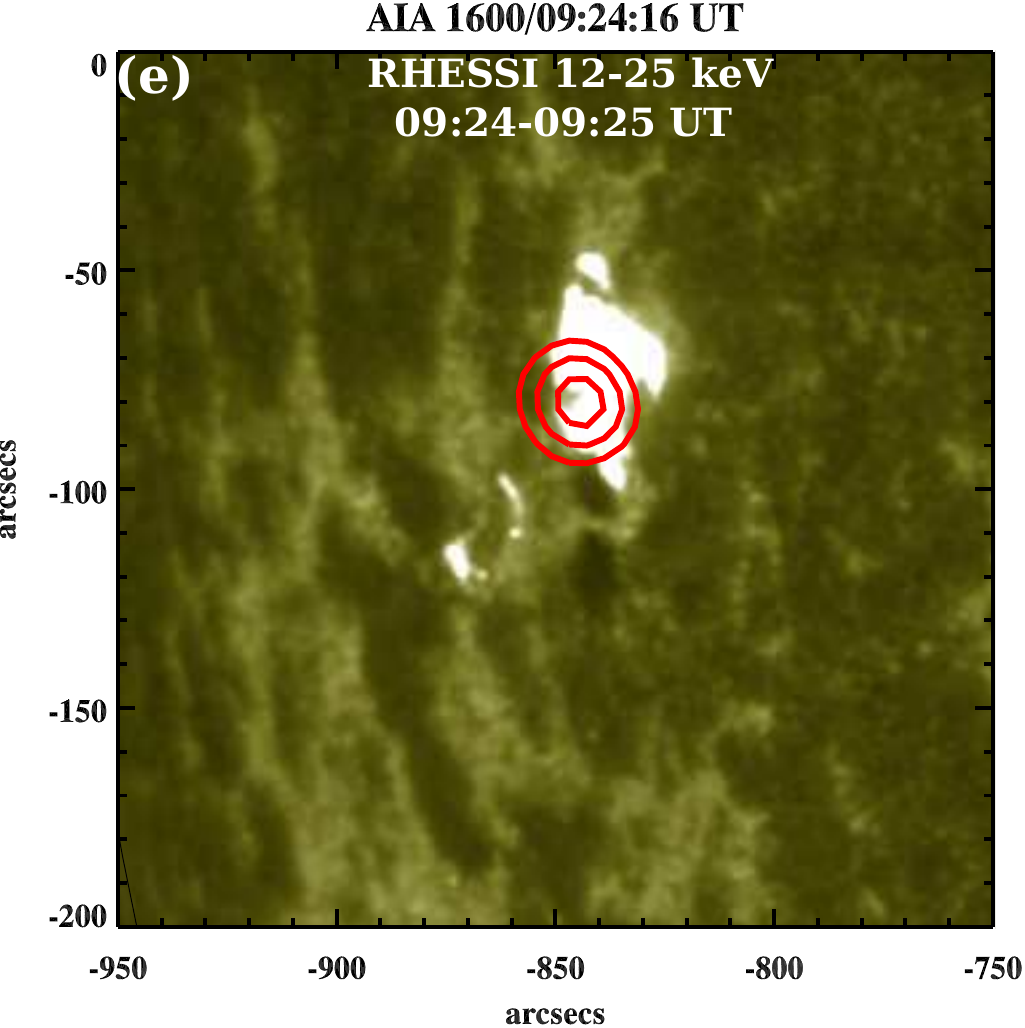}
\includegraphics[width=8.0cm]{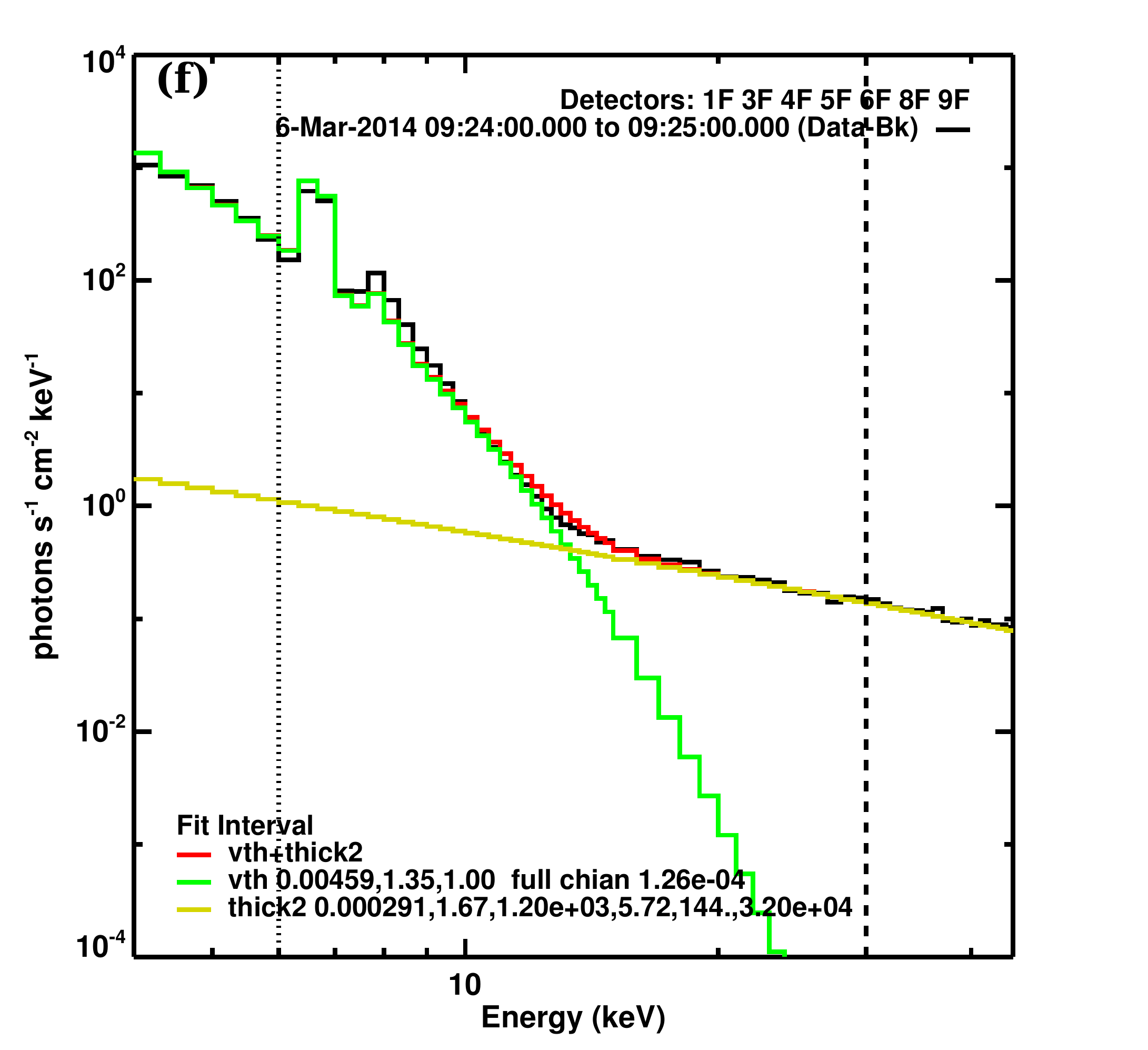}
}
\caption{(a,c,e) AIA 171 and 94 \AA~ running and base difference images during the propagating disturbance along the loop system. RHESSI hard X-ray source (12-25 keV) overplotted on the AIA 94 and 1600 \AA~ images (contour levels: 50$\%$, 70$\%$, and 90$\%$ of the peak intensity). (b,d) The temporal variation of the intensity along the selected path in 171 and 94 \AA~ images. (f) RHESSI X-ray spectrum fitted with an isothermal (green) and thick-target bremsstrahlung (yellow) components.  
(Animation is available.)}
\label{sl}
\end{figure*}

\subsection{Partial reflection and oscillation of hot loops}

The onset of the EUV wave and flare energy release
occurred simultaneously at $\sim$09:23~UT.
The partial reflection of the EUV wave was observed along the arcade
loops in the 171 \AA\ and 193 \AA\ channels.

Figures~\ref{sl}(a) and (c) show the running difference images in AIA 171 and
94 \AA~ channels, respectively. To investigate the wave propagation, we
selected a path along the loop (in blue color) and extracted the 171 and
94~\AA\ running-difference intensity ($\Delta$t=1 min) during
09:20--09:40~UT. Figure \ref{sl}(b) displays the stack plots of the intensity
distribution along the selected loop (shown by blue dots). The AIA 171~\AA\ stack
plot (panel b) reveals the propagating wave along the loop and partially
reflecting EUV wave from the opposite footpoint of the arcade loops (green
dashed line). The first two white arrows indicate the time of the EUV wave onset
and the time of its returning (after partial reflection)
(see 171 \AA~ movie). The travel
time is 10~min.

\begin{figure*}
\centering{
\includegraphics[width=12cm]{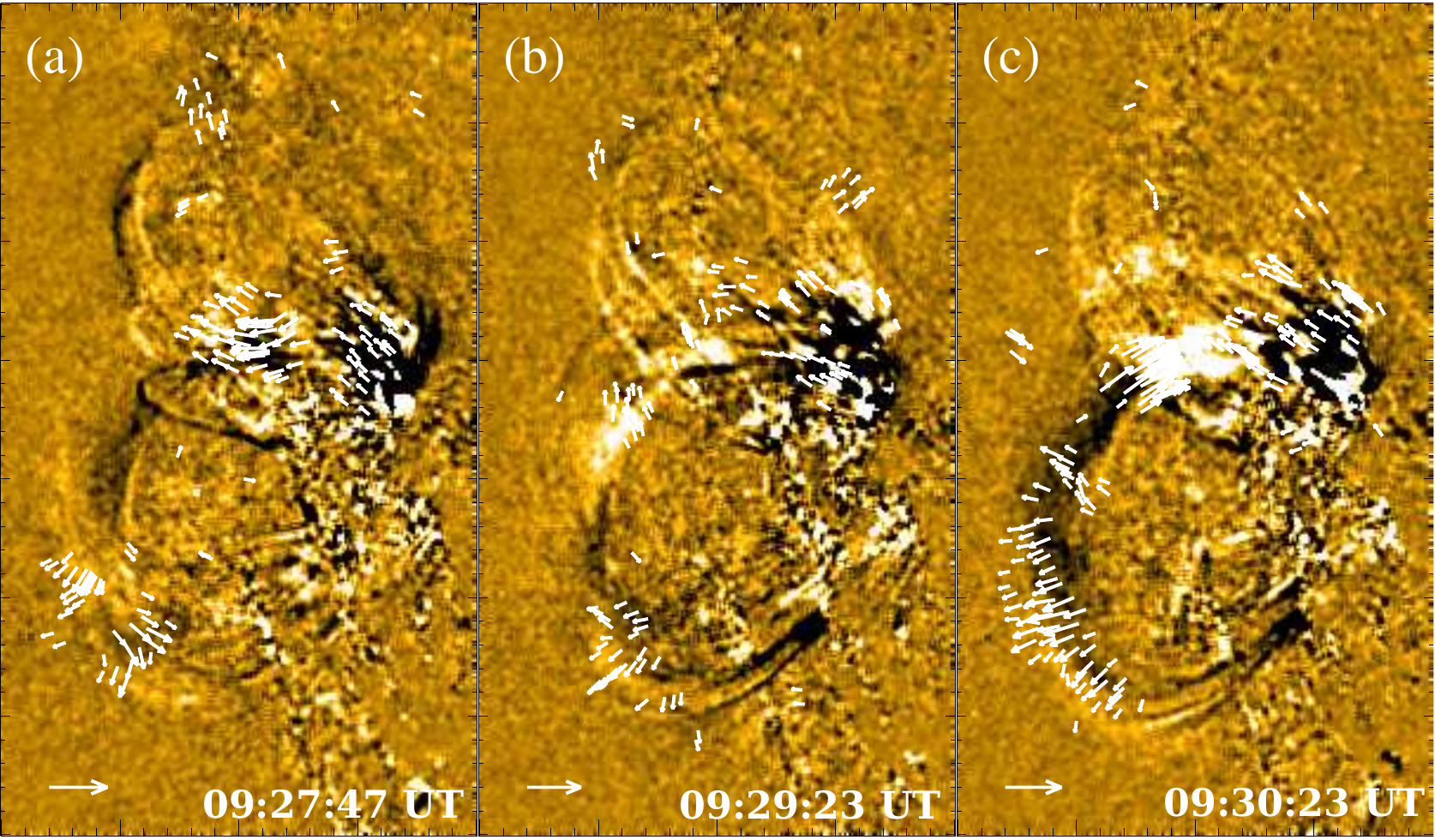}
}
\caption{Flows around the flare site at three different phases:
(a) outward propagation of the EUV wave; (b) backward propagation and onset of dimming;
(c) the loops' upward expansion and motion
back to the flare site. The length of the arrow in the bottom left represents
150~\kms.}
\label{flows}
\end{figure*}
\begin{figure*}
\centering{
\includegraphics[width=7cm]{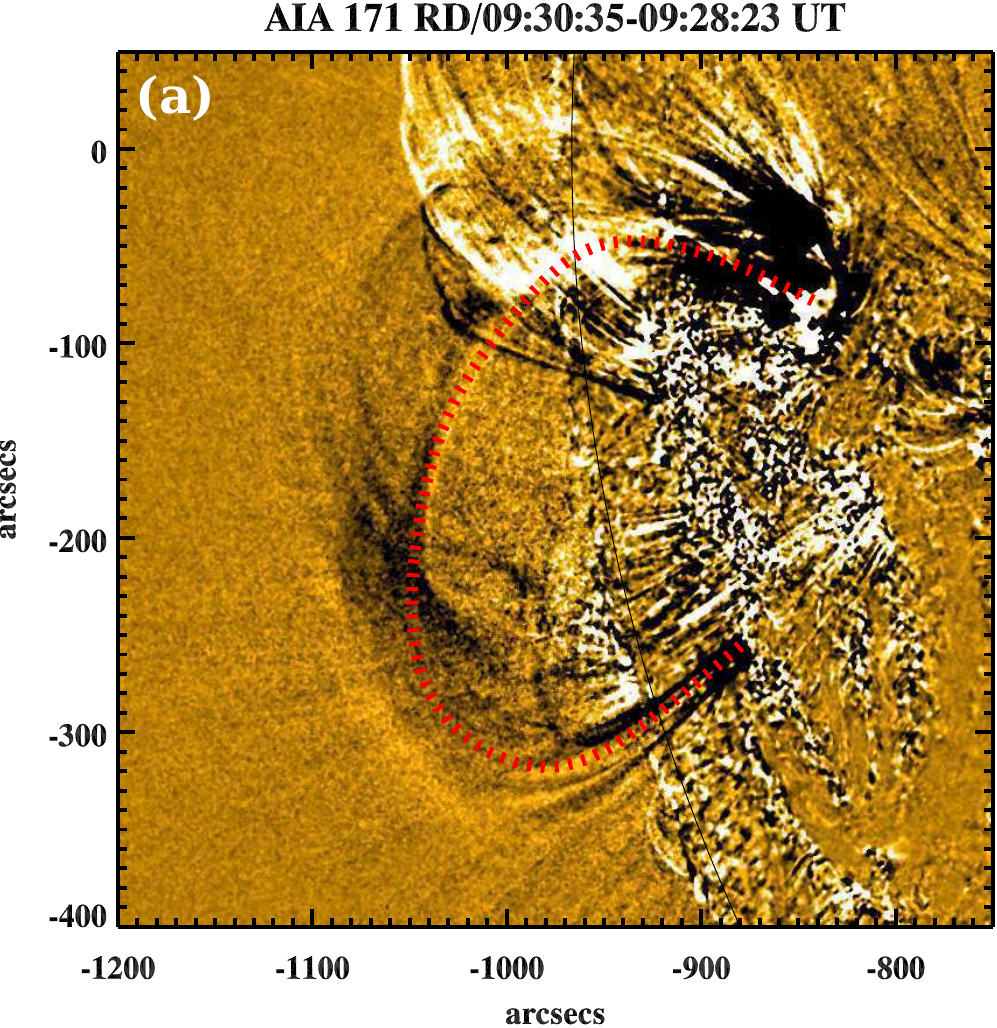}
\includegraphics[width=6cm]{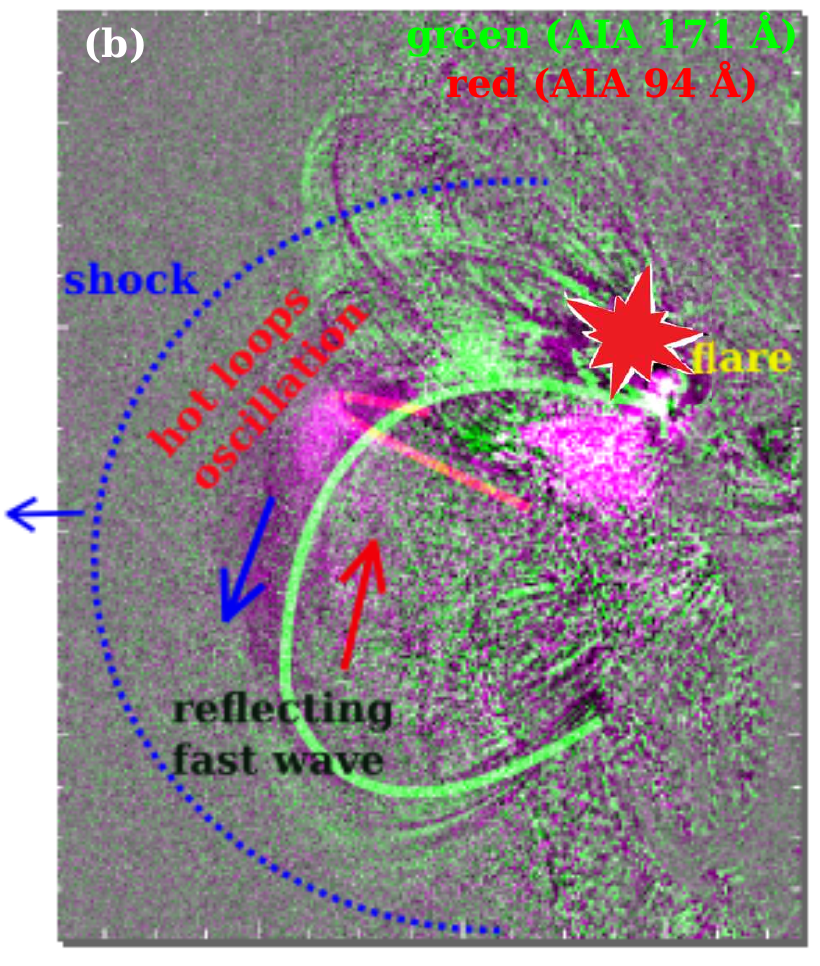}

\includegraphics[width=8.2cm]{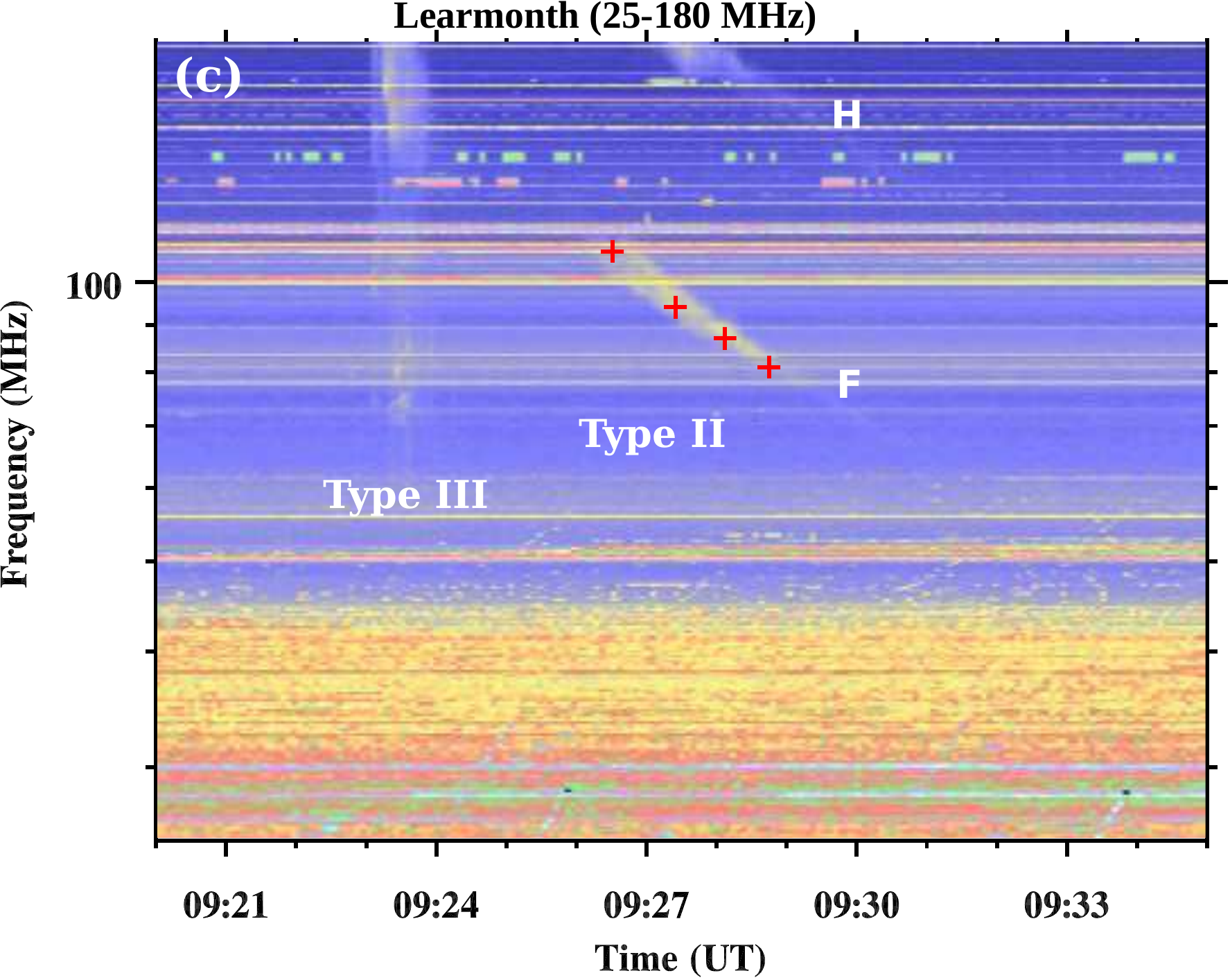}
\includegraphics[width=8.0cm]{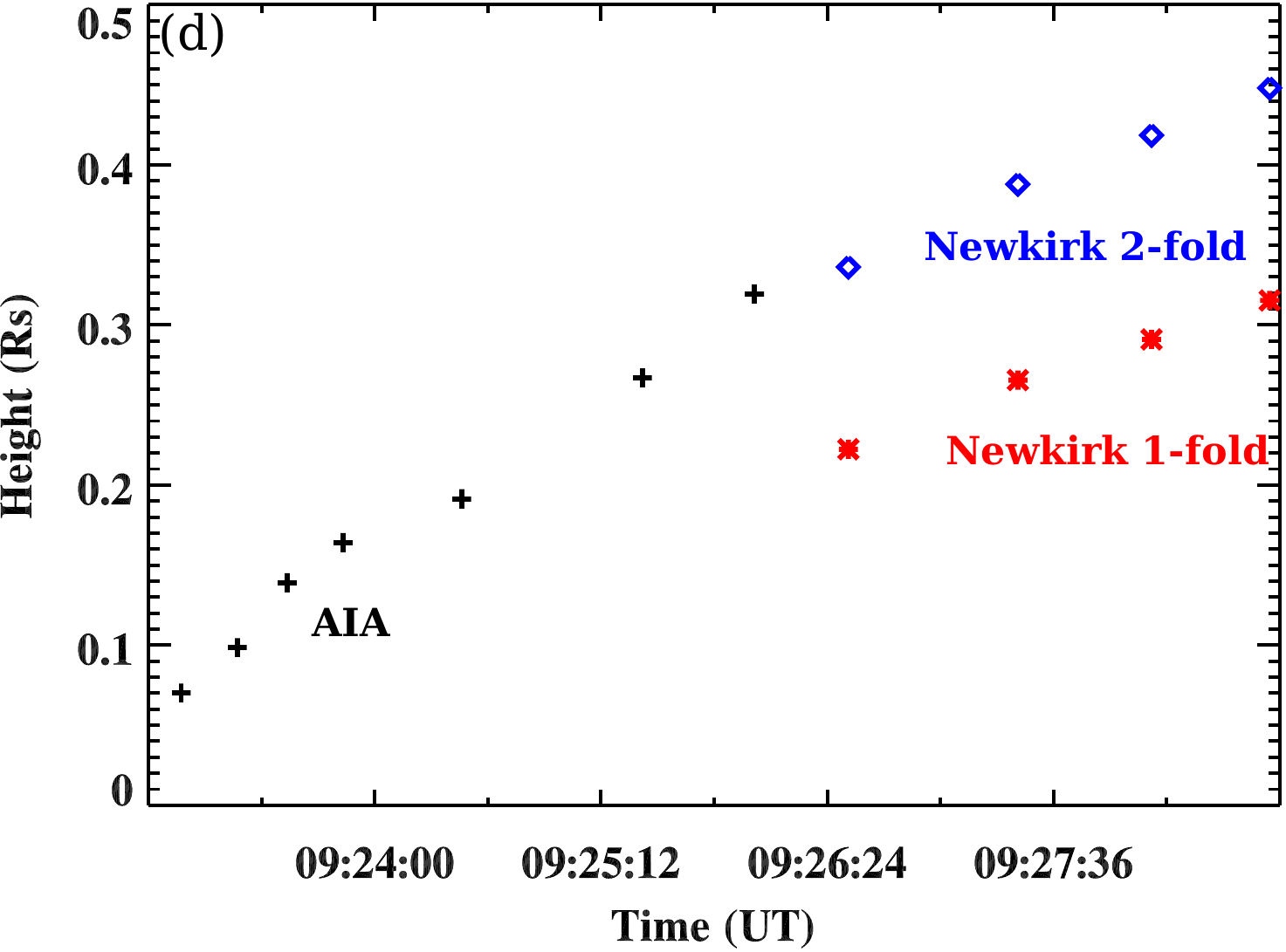}

}
\caption{(a) Loop length estimation by curvature radius maximization method
using AIA 171 \AA~ running difference image. (b) AIA 171 (green) and 94 (red) \AA~ blended image with schematic cartoon over it. (c) Learmonth dynamic radio spectrum. F and H are the fundamental and second harmonic bands of the type II radio burst. (d) Shock height-time plot from the AIA 171 \AA, Newkirk one-fold, and two-fold density models.
}
\label{fig5}
\end{figure*}

During its outward propagation, the EUV wave encountered an arcade of
loops overlying the 171~\AA\ loops.
These overlying loops became visible in the 94 and 131~\AA\ channels,
implying that they were heated during flare onset. One footpoint was
attached to the flare site (Figure~\ref{sl}(c)) so the initial heating was most likely caused
by energetic particles generated in the flare. We observed a single RHESSI \citep{lin2002} hard X-ray source (12-25 keV) during the flare maximum (09-24-09:25 UT). Simultaneously, we noticed impulsive footpoint heating/brightening in the AIA 1600 \AA~ channel (refer to the AIA 1600 \AA~ movie). Acceleration of nonthermal electrons possibly precipitated to the footpoint causing impulsive heating in the AIA 1600 \AA~ channel (Figure~\ref{sl}(e)). Figure ~\ref{sl}(f) shows the RHESSI spectrum fitted (energy range, 6-30 keV) with isothermal (T=15.6 MK, EM=4.5$\times$10$^{46}$ cm$^{-3}$) and thick-target bremsstrahlung (spectral index=5.7) components during 09:24-09:25 UT. We can see that the non-thermal contribution dominates over the thermal component above $\sim$13 keV. In addition, we observed a type III radio burst at 09:23-09:24 UT, suggesting the acceleration of non-thermal electrons upward into the interplanetary medium (Figure 5c). 
However, thermal contribution from the flare site may contribute to heating the loop in the later phase.

The heated loops are more visible in the AIA 94~\AA\
than in the AIA 131~\AA \ channel.
 The hot loops swayed in time with the
EUV wave seen along the 171~\AA\ loops.
Figure~\ref{sl}d shows a stack plot (base and running difference) of the 94~\AA\
emission along the selected slice in Figure~\ref{sl}c.
The transverse oscillation of the hot loops was triggered by the outward
propagating
fast-mode wave and partially reflected wave.
The arcade movement ($\sim$312 km s$^{-1}$) observed in the 94~\AA\ images is consistent with the
passage of the EUV wave observed in the 171~\AA\ running difference
images (Figure \ref{sl}b).

There is a dimming along the 171~\AA\ loops at about 09:30 UT
in the running difference movies (AIA 171 and 193 \AA\ channels).
The dimming is visible in Figure~\ref{sl}b. It occurred over a large region of
the loops, just after the partial reflection. At the same time a
171~\AA\ front was seen descending towards the flare footpoint.

To track the direction of the waves over the regions, we used the optical flow code of
\citet{gissot2007} on the series of 171~\AA\ images.
Figure~\ref{flows} shows  snapshots of the motions at three
times. In Figure~\ref{flows}a, one sees propagation up from the flare site,
a small region of outward motion to the north, as well as significant motion
south along the 171~\AA\ loops.
Figure~\ref{flows}b shows the motion soon after the partial reflection, at the time of the
onset of the dimming. Along the loop arcade, the motion is outward
from the site of the dimming: both back toward the flare site and
outward into the corona. The third image shows strong flows back to
the flare site as well as extensive expansion of the loop system
into the corona.
Typically the highest plane-of-sky velocities are $\sim$150~\kms\ (the length
of the arrows in the bottom-left corner).
 It is much less than the speed from the stack plot
which
measures the speed of faint fronts.

If we believe that the stack plot reveals
the propagation of the wave front, then
to deduce the phase speed of the wave,
we need to estimate the loop length. STEREO images could be
used to obtain the loop length with 3D reconstruction,
but the footpoints of the arcade loops were not observed by
STEREO because they were behind the western limb from STEREO.
We used the curvature radius maximization method \citep{asc2009} to estimate the
loop length.
We reconstructed the z axis using this method,
which provides a robust estimate of the 3D loop geometry from a single point
observation. Figure~\ref{fig5}a shows the best-fitted loop (red dotted curve).
The estimated loop length is $\sim$600\arcsec and the travel time was 10~min, so
the phase speed of the wave  $2L/P \sim$1450~km~s$^{-1}$.

{}

Figure \ref{fig5}b displays the blended AIA 171 (green) and 94 (red) \AA~ image, and
a schematic cartoon over the image summarizing the whole activity.
There was a partial AR filament eruption at $\sim$09:32~UT (marked by the
dotted line), just after the returning wave reached the flare site.
The AIA 171~\AA\ movie
shows a surge-like ejection of filament material from the flare site.
The EUV wave may
play an important role in the triggering of the cool plasma ejection.

Figure \ref{fig5}c shows radio dynamic spectrum (25-180 MHz) from Learmonth radio observatory. We selected data points (+) from the fundamental band (F) of type II radio burst to estimate the shock height using Newkirk one-fold and two-fold density model \citep{newkirk1961}. Figure \ref{fig5}d displays the shock height calculated from the AIA 171 \AA~ channel (+), Newkirk one-fold (blue star), and two-fold (red diamond) density models. Note that AIA height is basically the projected height (in the sky-plane) of the shock front from the flare center. The estimated shock speeds from the Newkirk one- and two-fold density models are 556-451 km s$^{-1}$ and 667-545 km s$^{-1}$, respectively. These speeds are consistent with the decelerating shock speed from AIA.  Moreover, the shock height is more consistent with the Newkirk two-fold density model.


\section{DISCUSSION AND CONCLUSION}
We report for the first time the propagation of a fast-mode wave along and
perpendicular to arcade loops, and its partial reflection from the
opposite footpoint of the arcade loops (in 171 \AA\ and 193 channels).
The phase speed of the longitudinal wave was $\sim$1450 km s$^{-1}$.

The EUV wave caused swaying motion of
an overlying arcade seen in AIA 94 and 131~\AA.
The initial plane-of-sky swaying velocity of the arcade was about
 312~\kms, and it was rapidly damped. Note that this speed is based on the
 initial amplitude of the hot loop oscillations, and is not the phase speed.
  Assuming the arcade consists of
 semi-circular loops with height $\sim$200$\arcsec$ (Figure~\ref{sl}(c)), and a phase speed of the order of Alfv\'en speed,
 $\sim$1900~\kms, the
 kink mode would have an oscillation period of about 8~min which is
 roughly the back-and-forth period observed in Figure~\ref{sl}(d).


The nature of the observed fast MHD wave in closed arcade loops
may be similar to the propagating fast-mode wave trains observed in the open-fan loops \citep{liu2011}. Quasiperiodic wave trains are best observed in the AIA 171 \AA~ channel \citep{liu2011,ofman2011,kumar2013a,pascoe2013}. However, here we observed only a single front (not wave trains) in the arcade loops and its close association with a flare. Theoretically, fast-mode waves can be guided by the magnetic field and trapped in a region of low Alfv\'enic speed (i.e., high density loops) \citep{roberts1984,nakariakov2005}.

 Quasi-periodic wave trains are associated with quasi-periodicity of the
 flare (e.g., periodic reconnection). These signatures are generally observed
 in the hard X-ray or AIA 1600 \AA~ channels (as a result of periodic particle
 acceleration). In our case, we do not see any quasi-periodic pulsation in the AIA 1600 \AA~ or hard X-ray channels.
 Therefore, this event is related to a single burst of energy release that generates a
 single wave front.

The dome like expansion of a fast-mode wave was also observed in the AIA 193
and 211 \AA~ channels. The
lateral expansion (in the southward direction) of the dome-shaped wave
coexists with the EUV wavefront propagating along the loops.
 The fast-mode wave propagates
perpendicular and parallel to the magnetic field. Theoretically, the fast MHD
wave can transport its energy in any direction (i.e., parallel or
perpendicular to the magnetic field).

 The fast-mode wave was closely associated with the compact flare.
 The EUV wave might be triggered either
by the impulsive expansion of the flare loop or by thermal pressure (blast)
generated during the flare energy release \citep{vrsnak2008,kumar2013c}.
 In our case, we do
not observe impulsive plasmoid ejection in either the cool or hot AIA channels
 before or during the onset
of the EUV wave.
Therefore, it is unlikely that the EUV wave observed here was triggered by the plasmoid ejection as reported in previous
case studies \citep{klein1999,kumar2013a}. There was no CME loop behind the
EUV wave as observed in \citet{veronig2010}. Also we do not see two wave
components (i.e., slow and fast), where the speed of the fast-mode wave (true
shock wave) should be almost three times that of the slow wave (pseudo wave)
\citep{chen2002,chen2011,kumar2013b}. The wave is most likely generated
impulsively by the flare-related pressure pulse (or blast wave), and then
propagated freely  in the ambient medium. A metric type II radio burst also reveals the presence
of a shock wave in the corona.

In conclusion, we presented the first observation of partial reflection and trapping of a
fast magnetoacoustic wave in coronal arcade loops. Reflection and
transmission of shock waves from the boundary of a coronal hole
\citep{gopal2009,olmedo2012} or from a different active region
\citep{kumar2013a} have been observed previously but, so far, there is no
observational report of the reflecting and trapping of fast-mode waves in coronal arcade
loops. When the fast-mode wave encounters a region of large gradient (from low to high) in the Alfv\'en speed, they may experience partial reflection. Thus a fast-mode wave traveling in a loop may be internally reflected at the loop edge and stay trapped in the loop.
This observation also has an implication on the origin of the shock
wave in the context of flares. We agree that most of the low coronal
shocks are usually driven by the CME piston. However, here the observation of
a reflecting fast-mode wave in coronal loops strongly favours the flare origin
(in this event) for the generation of the low coronal shocks rather than
CMEs.
Multiwavelength studies of similar
 events will be helpful to understand the excitation mechanisms of these waves in more detail.

\acknowledgments
We thank the referee for the positive and constructive comments/suggestions that improved the manuscript considerably.
SDO is a mission for NASA's Living With a Star (LWS) program.  RHESSI is a NASA Small Explorer. We would like to thank Prof. Valery M. Nakariakov, Hardi Peter, and Robert Cameron for several fruitful discussions.

\bibliographystyle{apj}


\end{document}